\documentclass[%
superscriptaddress,
preprint,
 amsmath,amssymb,
 aps,
]{revtex4-1}

\usepackage{multirow,makecell}
\usepackage[table]{xcolor}
\usepackage{mathptmx}
\usepackage{graphicx}
\usepackage{dcolumn}
\usepackage{bm}
\usepackage[normalem]{ulem}  
\usepackage{todonotes}

\usepackage{hyperref}
\hypersetup{
    colorlinks=true,
    linkcolor=blue,
    filecolor=blue,      
    urlcolor=blue,
    citecolor=blue,
}
\newcommand{\mudpb}{\mu_{\rm d}^{\rm pb}}

\newcommand{\muspb}{\mu_{\rm s}^{\rm pb}}
\newcommand{\Fspb}{F_{\rm s}^{\rm pb}}

\newcommand{\mus}{\mu_{\rm s}}
\newcommand{\mud}{\mu_{\rm d}}

\begin{document}

\preprint{APS/123-QED}

\title{Intruder in a two-dimensional granular system: Effects of dynamic and static basal friction on stick-slip and clogging dynamics}

\author{C. Manuel Carlevaro}
\affiliation{Instituto de F\'isica de L\'iquidos y Sistemas Biol\'ogicos, CONICET, 59 789, 1900 La Plata, Argentina}
\affiliation{Dpto. Ing. Mec\'anica, Universidad Tecnol\'ogica Nacional, Facultad Regional La Plata, Av. 60 Esq. 124, La Plata, 1900, Argentina}

\author{Ryan Kozlowski}
\affiliation{Department of Physics, Duke University, Durham, North Carolina 27708, USA}

\author{Luis A. Pugnaloni} 
\affiliation{Dpto. de F\'isica, Fac. Ciencias Exactas y Naturales, Universidad Nacional de La Pampa, CONICET, Uruguay 151, 6300 Santa Rosa (La Pampa), Argentina}

\author{Hu Zheng}
\affiliation{Department of Physics, Duke University, Durham, North Carolina 27708, USA}%
\affiliation{Department of Geotechnical Engineering, College of Civil Engineering, Tongji University, Shanghai, 200092, China}

\author{Joshua E.~S.~Socolar}
\affiliation{Department of Physics, Duke University, Durham, North Carolina 27708, USA}

\author{Lou Kondic}
\affiliation{%
 Department of Mathematical Sciences and Center for Applied Mathematics and Statistics, New Jersey Institute of Technology, Newark, New Jersey 07102, USA
}%



\date{\today}

\begin{abstract}

We discuss the results of simulations of an intruder pulled through a two-dimensional granular system by a spring, using a model designed to lend insight into the experimental findings described by Kozlowski et al. [Phys. Rev. E {\bf 100}, 032905 (2019)].  In that previous study the presence of basal friction between the grains and the base was observed to change the intruder dynamics from clogging to stick--slip. Here we first show that our simulation results are in excellent agreement with the experimental data for a variety of experimentally accessible friction coefficients governing interactions of particles with each other and with boundaries. Then, we use simulations to explore a broader range of parameter space, focusing on the friction  between the particles and the base.  We consider a range of both static and dynamic basal friction coefficients, which are difficult to vary smoothly in experiments.  The simulations show that dynamic friction strongly affects the stick--slip behaviour when the coefficient is decreased below 0.1, while static friction plays only a marginal role in the intruder dynamics.

\end{abstract}

\maketitle

\section{Introduction}

Granular media respond in a variety of ways to applied loads, such as boundary shear \cite{granularfaultdanielshayman,localglobalavalanchesbares,SimGranularSeismicPicaCiamarra2011}, intruding rods \cite{stickslipalbert,stickslippulldiskoutMetayer}, or surface sliders \cite{stickslipnasuno,stickslipcracklingagheal,SimGranularSeismicPicaCiamarra2011}, exhibiting behaviors that include fluid-like flow, solid-like rigidity, and (sometimes periodic) cycles of stability and failure \cite{jaeger96b}. The response of a granular medium to a point-load, or single-grain perturbation, is a particularly sensitive probe of the connection between grain-scale dynamics and large-scale material stability and failure. Driving with a single particle avoids averaging over interactions of many grains with the extended objects (system boundaries, sliders, etc.) that are typically used to apply stress. Recent granular point-load studies have focused on the dynamical response of a single-grain intruder, mostly in quasi-two-dimensional (2D) beds of disks \cite{probeintruderreichhardt,intrudervibrationgdauchot,dragforcecavityformationintruderkolb,slowdraggeng,rheologyintruderexperimentseguin,sticksliptordesillas}. In all cases, the dynamics is affected by the packing fraction and the strength of the driving mechanism, whether the intruder is driven at a constant velocity, by a constant force, or by a continually loading spring. An experiment on intruder dynamics in a 2D Couette geometry showed that friction between the particles and the supporting substrate (basal friction) also has a strong effect \cite{Kozlowski2019}, as had previously been observed only in the context of quasistatic shear-jamming \cite{shearjamnobfhu}.

In the present work, we develop numerical simulations and validate them against the experimental findings in ~\cite{Kozlowski2019} with the aim of elucidating the influence of basal friction on the dynamics of an intruder moving through a granular medium, which is important for understanding the role of basal friction in supporting force-bearing clusters of grains subjected to a point load. We use numerical simulations to extend our study to parameter regimes that are difficult to access in the laboratory.  In particular, we smoothly vary the basal dynamic and static friction coefficients from zero to the experimentally relevant value.  We find, on the one hand, that dynamic friction $\mus$ controls the overall dynamics with a clear stick--slip behavior for $\mus > 0.1$ and an intermittent clogging-like flow for $\mus < 0.1$. Static friction, on the other hand, plays a marginal role in determining the intruder's behavior.

The paper is organized as follows. In Sec.~\ref{sec:exp}, we review the experimental setup, and in Sec.~\ref{sec:simulations}, we describe the simulation procedures. Section~\ref{sec:results} contains our key results.  After describing the basic dynamics of the  intruder,
we compare the experimental and simulation results for different packing fractions 
for both frictional and frictionless substrates. 
We then present simulation results for a range of static and dynamic basal friction coefficients. 
Section \ref{sec:conclusions} is devoted to our conclusions regarding the effect of basal friction on the intruder dynamics. 

\section{The Experiment}\label{sec:exp}

\begin{figure}
    \centering
    \includegraphics[width=0.5\columnwidth]{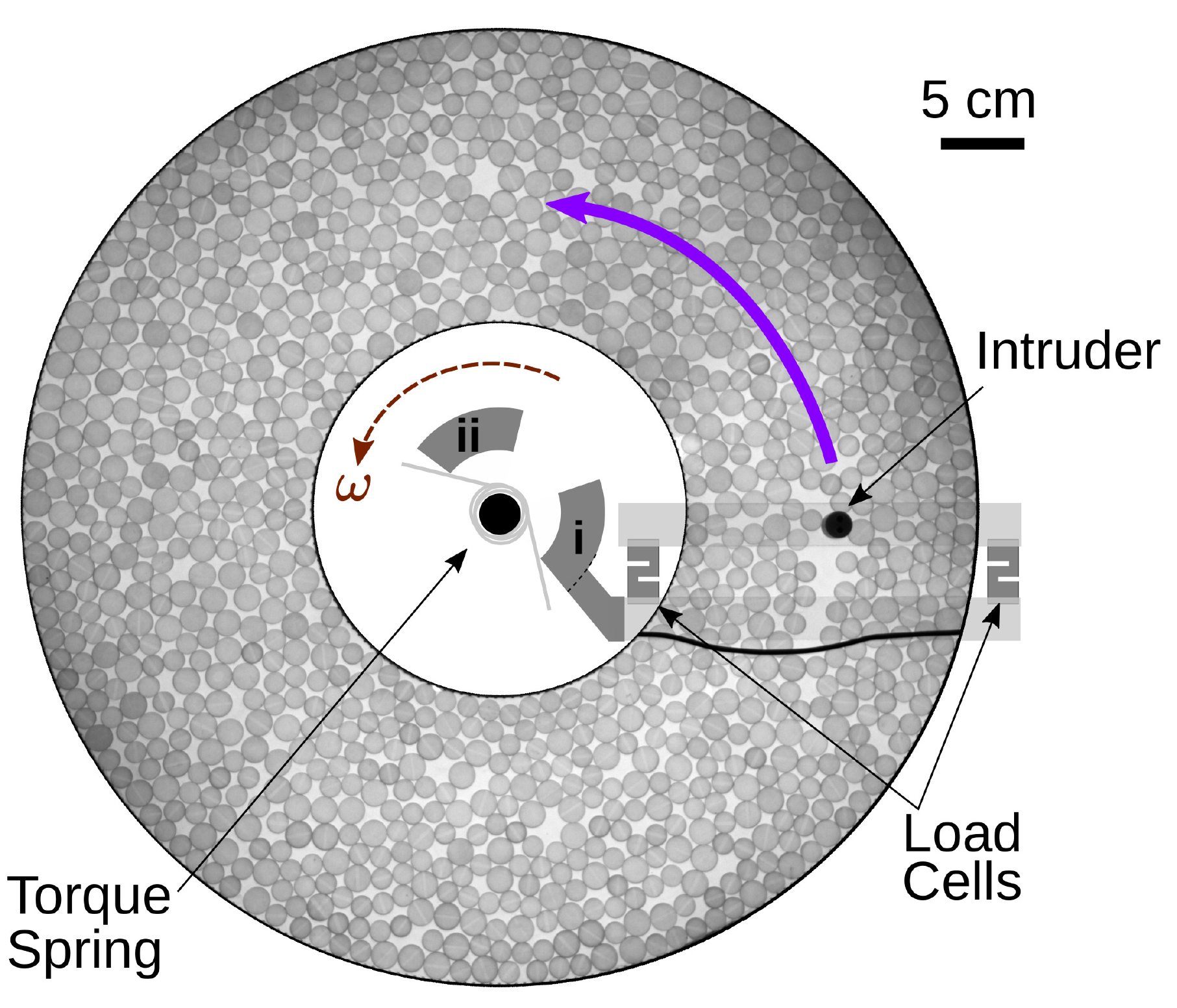}
    \caption{A top-down schematic of the experiment with a sample image of grains.
    } 
    \label{fig:experimentapp}
\end{figure}

The experimental system consists of a layer of bidisperse plastic disks confined to an annular region by solid walls.  An intruder the size of one disk is pushed in the azimuthal direction around the annulus by means of an externally controlled arm attached to a torsion spring whose other end is driven at constant angular speed. 
The setup, shown in Fig.~\ref{fig:experimentapp}, and observations have been described in detail in Ref.~\cite{Kozlowski2019}; here we provide a brief summary. 

A central axle supports a pusher arm rigidly attached to an intruder rod suspended in a bidisperse granular bed (grain diameters $d_{\rm s} = 1.28\,$cm and $d_{\rm l} = 1.60\,$cm). The grains are confined in an annular Couette geometry having a channel width of $17.8$ cm, or roughly 14 small particle diameters, and boundaries lined with ribbed rubber to minimize boundary slipping. The central axis is composed of a bottom rotating shaft (indicated as {\bf ii} in Fig.~\ref{fig:experimentapp}) driven by a stepper motor at constant angular velocity $\omega = 0.119 \pm 0.006$ rad/s, and a top rotating shaft (indicated as {\bf i} in Fig.~\ref{fig:experimentapp}) that is coupled to the bottom shaft by a torque spring ($\kappa = 0.431 \pm 0.001$ Nm/rad). The intruder (diameter $d_I = 1.59 \pm 0.01\,$cm, the size of a large grain) is held at a fixed radius from the annulus center ($R = 19.7 \pm 0.1$ cm). Load cells in the pusher arm measure the force between the granular system and the intruder.   The torque spring loads the system until the granular medium yields and the intruder slips. Cameras above the annular bed allow for intruder and particle tracking. The force on the load cells is recorded every $0.01\,$s, and the cameras capture images every $0.02\,$s. 

For experiments with basal friction, particles are dry and lightly coated in powder to reduce friction.  The static friction coefficient between the particles and the glass table was measured to be $\sim 0.36$, and we assume that the dynamic friction coefficient is slightly below this value. To perform experiments without basal friction, the annular region is filled with water, and particles float on water.  In this case, the particle diameters are $d_{\rm s} = 1.30\,$cm and $d_{\rm l} = 1.62\,$cm, slightly larger than the dry case due to swelling of the plastic. In both experiments, the number ratio of large to small particles is approximately fixed at $1\,:\,2.75$. 

\section{Simulations}\label{sec:simulations}

We have carried out discrete element method (DEM) simulations of two-dimensional (2D) systems of particles. The simulations were implemented by means of the Box2D library~\cite{box2d}, which uses a constraint solver to handle rigid bodies. Before each time step, a series of iterations (typically 100) is used to resolve constraints on overlaps and on static friction between bodies through a Lagrange multiplier scheme~\cite{catto}. After resolving overlaps, the inelastic collision at each contact is solved and new linear and angular velocities are assigned to each body. The equations of motion are integrated through a symplectic Euler algorithm. Solid friction between grains is also handled by means of a Lagrange multiplier scheme that implements the Coulomb criterion with the dynamic and static friction coefficients set to be equal. The approach yields realistic dynamics for granular bodies \cite{pytlos2015modelling} with complex shapes, including sharp corners, and has been successfully used to study grains under tapping protocols~\cite{carlevaro_jsm11,irastorza2013exact} and under vigorous vibration \cite{sanchez2014effect}.

\begin{figure}
    \centering
    \includegraphics[width=0.5\columnwidth]{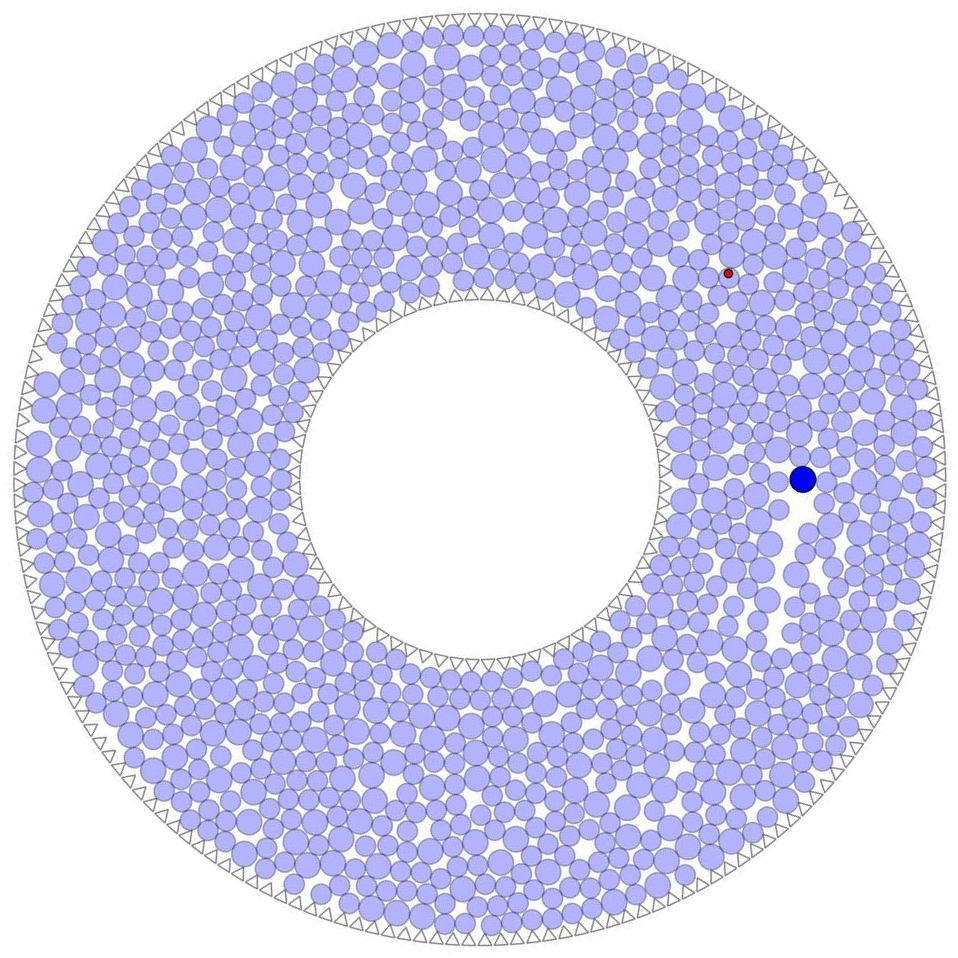}
    \caption{Snapshot of a sample simulation. The light blue particles are bidisperse (see text for details), and the static inner and outer boundaries are formed by equilateral triangles. The blue disk is the intruder (which is moving counter-clockwise around the annulus). The red dot indicates the position of the end of the torque spring that is driven at constant velocity; the other end of the spring is attached to the blue particle.
    A movie of the dynamics is available as Supplementary Material.  
    } 
    \label{fig:setup}
\end{figure}

\subsection{Cell}

Figure \ref{fig:setup} shows a snapshot of a typical simulation. The cell consists of two immobile concentric boundary ``rings'' forming an annular two-dimensional (2D) Couette cell. The boundaries are formed by small equilateral triangles facing inward (toward the annular channel), which prevent slippage at the boundary, serving the role of the ribbed rubber in the experiments. The inner ring consists of 72 triangles of side length $0.682 d$, where $d$ is the diameter of the small particles used for the granular pack (see below); the outer ring consists of 180 triangles of side length $0.683 d$. The inner and outer rings are $8.810 d$ and $22.800 d$ in radius, 

Gravity (with acceleration $g$) acts in the direction perpendicular to the Couette plane. The base on which the circular particles rest is modeled by implementing an effective solid friction as follows. If a particle is moving at a speed above a small threshold $v'$ (i.e., $|\bm{v}|>v'$), then a dynamic friction force with the base $F^{\rm pb}_{\rm d} = -\mudpb mg \bm{v}/|\bm{v}|$ is applied to the center of mass of the particle. Whenever $|\bm{v}|<v'$, the particle is immobilized by setting $\bm{v}=0$. If at rest, the particle will only resume translational motion if the total external force exerted by other particles exceeds the static friction force with the base set to $\Fspb=\muspb mg$, with $\muspb=v'/(g \:dt)$, where $dt$ is the simulation time step. This ensures that a particle will resume motion only if its initial velocity due to the collisions in the previous time step exceeds the velocity threshold $v'$. Therefore, the static friction is controlled via $v'$. We do not implement rotational friction forces between the particles and the base in this model.

\subsection{Disks}

The annulus is filled with a large to small $1:2.75$ binary mixture of circular particles. The small particles have diameter $d$ and mass $m$. The larger particles have diameter $d_{\rm l}=1.25d$ and $m_{\rm l}=(d_{\rm l}/d)^2 m =1.565m$. The packing fraction is set by inserting a given number of disks ($N=N_{\rm s}+N_{\rm l}$, where $N_{\rm s}$ and $N_{\rm l}$ are the number of small and large particles, respectively), conserving the $1:2.75$ number ratio. Table \ref{tab:packing-fractions} lists all packing fractions explored. For the calculation of the packing fraction we 
exclude both the area of the triangles and the space between them. 

\begin{table}
    \centering
    \begin{tabular}{l r r}
         \hline\hline
         $\phi$ & $N_{\rm l}$ & $N_{\rm s}$\\
         \hline
    0.6480 & 245 & 676 \\     
    0.6589 & 250 & 685 \\     
    0.6691 & 253 & 697 \\     
    0.6797 & 257 & 708 \\  
    0.6899 & 260 & 720 \\     
    0.6998 & 264 & 730 \\     
    0.7104 & 268 & 741 \\     
    0.7203 & 272 & 751 \\     
    0.7308 & 276 & 762 \\     
    0.7414 & 280 & 773 \\     
    0.7513 & 284 & 783 \\     
    0.7619 & 288 & 794 \\     
    0.7724 & 292 & 805 \\     
    0.7824 & 296 & 815 \\     
         \hline\hline
    \end{tabular}
    \caption{List of packing fractions explored in the simulations and the corresponding numbers of large and small particles.}
    \label{tab:packing-fractions}
\end{table}

Each disk interacts with other disks, the boundaries of the cell, and the intruder disk (see below) as a perfectly rigid impenetrable object. The result of a collision is controlled by a restitution coefficient $\epsilon$ and the static $\mus$ and dynamic $\mud$ friction coefficients. Unless explicitly noted, we take $\mus=\mud$. Note that the particle--base interaction has $\muspb\neq\mudpb$. Table \ref{tab:interactions} lists the parameters for the various pair-wise interactions in the system.

\begin{table}
    \centering
    \begin{tabular}{r c r r r}
    \hline\hline
         & Acronym & $\mus$ & $\mud$ & $\epsilon$ \\
         \hline
        particle--particle & pp & 1.20 & 1.20 & 0.05 \\
        particle--annulus & pa & 0.77 & 0.77 & 0.05 \\
        particle--intruder & pi & 0.41 & 0.41 & 0.05\\
        particle--base & pb & [0.36;1.00] & [0;0.36] & -- \\
        intruder--base & ib & 0.00 & 0.00 & --\\
    \hline\hline    
    \end{tabular}
    \caption{List of friction and restitution coefficients for the various pair-wise interactions in the system.  The numerical values for the parameters are motivated by the ones measured in the experiments.}
    \label{tab:interactions}
\end{table}

\subsection{Intruder}

A circular particle of diameter $1.25 d$ (the size of a large particle) is used as an intruder. The intruder interacts only with the other disks, not with the base (see Table \ref{tab:interactions}) as in the experiments, where the intruder is suspended above the base at all times. 

The intruder is constrained to move along a circle of radius $15.8 d$ centered at the annulus center. This is done by binding the intruder to a very stiff radial spring. A ``soft'' torque spring ($K=3591.98 mgd/{\rm rad}$) is connected to the intruder and rotated counter-clockwise at a low constant angular speed, $\omega = 0.00432\sqrt{g/d}$. This drives the intruder through the pack of disks. The attached spring  can only pull the intruder; if the spring becomes shorter than its equilibrium length, no force is applied. The mass of the intruder is set to $380m$, a mass that yields the same moment of inertia with respect to the center of the annulus as the relevant moment of inertia in the experiment, which
includes both the pushing arm and the intruder. The simulation time step is $0.001\sqrt{d/g}$ and the instantaneous intruder position and velocity (and spring force) are recorded every 100 time steps. The time step is sufficiently small to avoid numerical instabilities, and the results are consistent for smaller time steps.


\section{Results}\label{sec:results}

For both experiments and simulations, the intruder is driven completely around the annulus at least twice and at most ten times in a given run. For all statistical analyses, we ignore the first revolution, in which transient effects are observed as the intruder moves through an initially random configuration of grains. The friction coefficients and packing fractions of the experiments, as well as the post-processing performed on intruder velocity and force data, are matched in simulation. We first validate simulations by comparing the statistics of intruder velocity, spring force, creep velocity during sticking periods and waiting times between them with those of the experiment, 
and then continue to explore the influence of basal friction.  

\subsection{Stick-slip dynamics \label{sec:stickslip}}

\begin{figure}
    \centering
    \includegraphics[width=1\columnwidth]{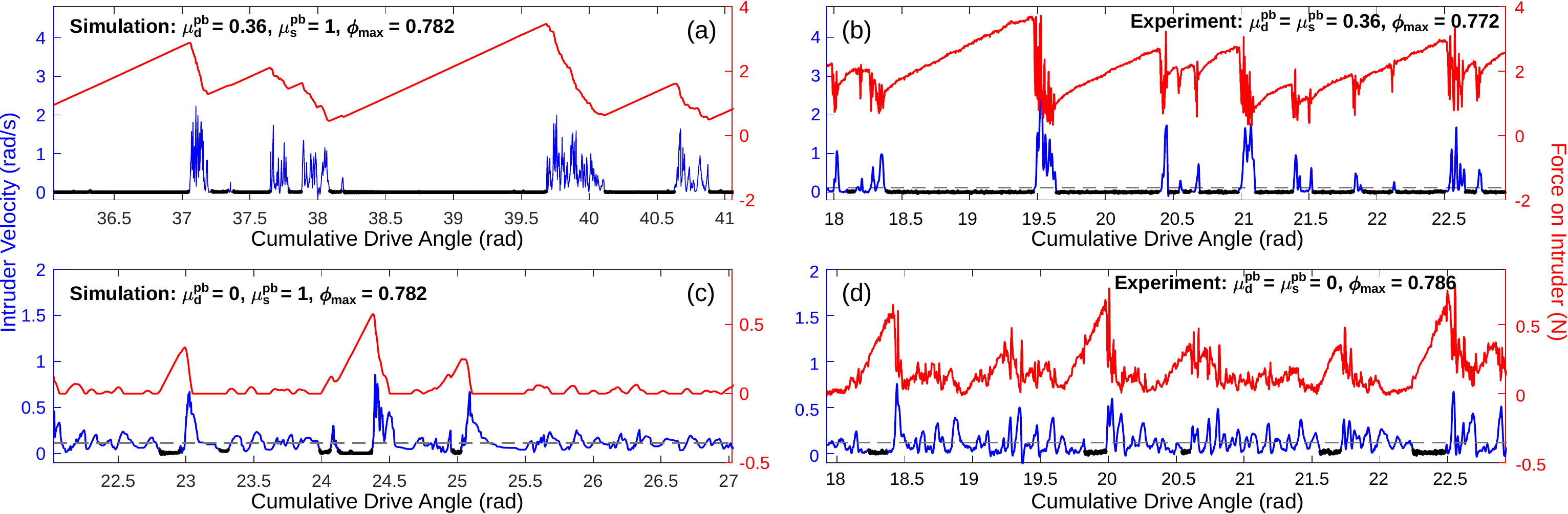}
    \caption{Sample time series at maximum packing fraction for cases with basal friction (a,b) and without (c,d) in both simulation (a,c) and experiment (b,d). Thick black overlay plots indicate detected sticking periods. 
    }
    \label{fig:comp-timeseries}
\end{figure}

Figure \ref{fig:comp-timeseries} shows examples of time series for the intruder velocity and force, where time is measured in terms of the cumulative drive angle $\theta = \omega t$. The force measured in experiments is the force of the grains acting on the intruder (compression of the load cells that hold the intruder) while in simulations the force presented is that exerted by the torque spring on the intruder. In a static configuration, these forces are the same, but while the intruder is moving, as during slips, the experimental force fluctuates more, as discussed further below. 
The packing fractions used for
Fig.~\ref{fig:comp-timeseries} are the highest packing fractions $\phi_{\rm max}$ explored in both simulation and experiment (which are slightly different in the simulations for different values of the basal friction coefficient). Higher packing fractions could not be studied with the present experimental apparatus due to buckling of the particles out of plane. The thick, black overlay plots on intruder velocity are detected sticking periods, which are defined as series of consecutive data points in the intruder velocity that fall below threshold 0.04 rad/s for a duration of at least 0.4 s. 
 
Figure~\ref{fig:comp-timeseries} shows that simulations (a, c) and experiments (b, d) at comparable packing fractions produce qualitatively similar results for two different values of the basal friction coefficient. For a frictional base ($\mudpb=0.36$), the intruder displays a clear stick--slip dynamics, characterized by extended sticking periods followed by rapid slip events. During sticking periods the intruder velocity is nominally zero and the force of the grains (and torque spring) acting on the intruder increases approximately linearly with time.  The granular medium eventually yields under the increasing point load, and a slip event occurs.  During a slip, the intruder's velocity fluctuates irregularly as it collides with many grains until the medium forms a stable structure again. In the experiments, the measured force fluctuates rapidly as the load cells register numerous collisions with grains. In the simulations, the measured force is of the torque spring acting on the intruder, with fluctuations that are relatively small compared to the force itself. 

For a frictionless base, both experiments and simulations show long periods of time during which the intruder moves at the drive speed with superimposed fluctuations. Occasionally, the intruder does get stuck for a short period and then slips. This behavior is reminiscent of clogging of grains that flow through a restricted aperture; we refer to it as intermittent flow dynamics \cite{Kozlowski2019}. 

\subsection{Comparison between simulations and experiments}\label{sec:comparison}

\begin{figure}
    \centering
    \includegraphics[width=1\columnwidth]{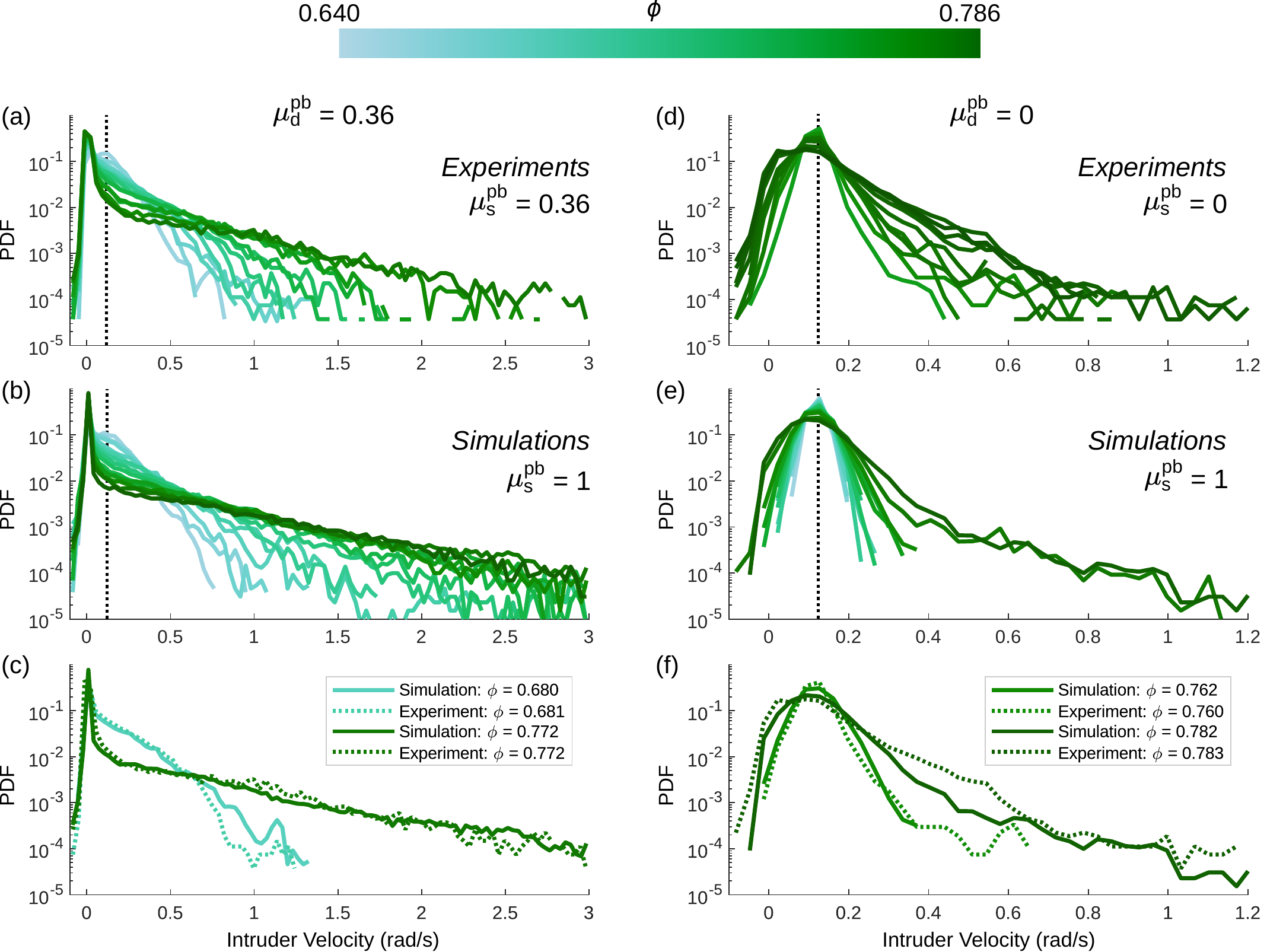}
    \caption{Intruder velocity distribution for frictional (a-c) and frictionless (d-f) bases. A range of packing fractions have been studied as indicated by the color scale. Experimental (a,d) and simulation (b, e) results show the same trends and range of velocities. Panels  (c) and (f) show a direct comparison of the experimental and simulation PDF for two specific values of $\phi$. The vertical dotted line in (a,b,d,e) marks the drive velocity.}
    \label{fig:pdf-velocity}
\end{figure}

\begin{figure}
    \centering
    \includegraphics[width=1\columnwidth]{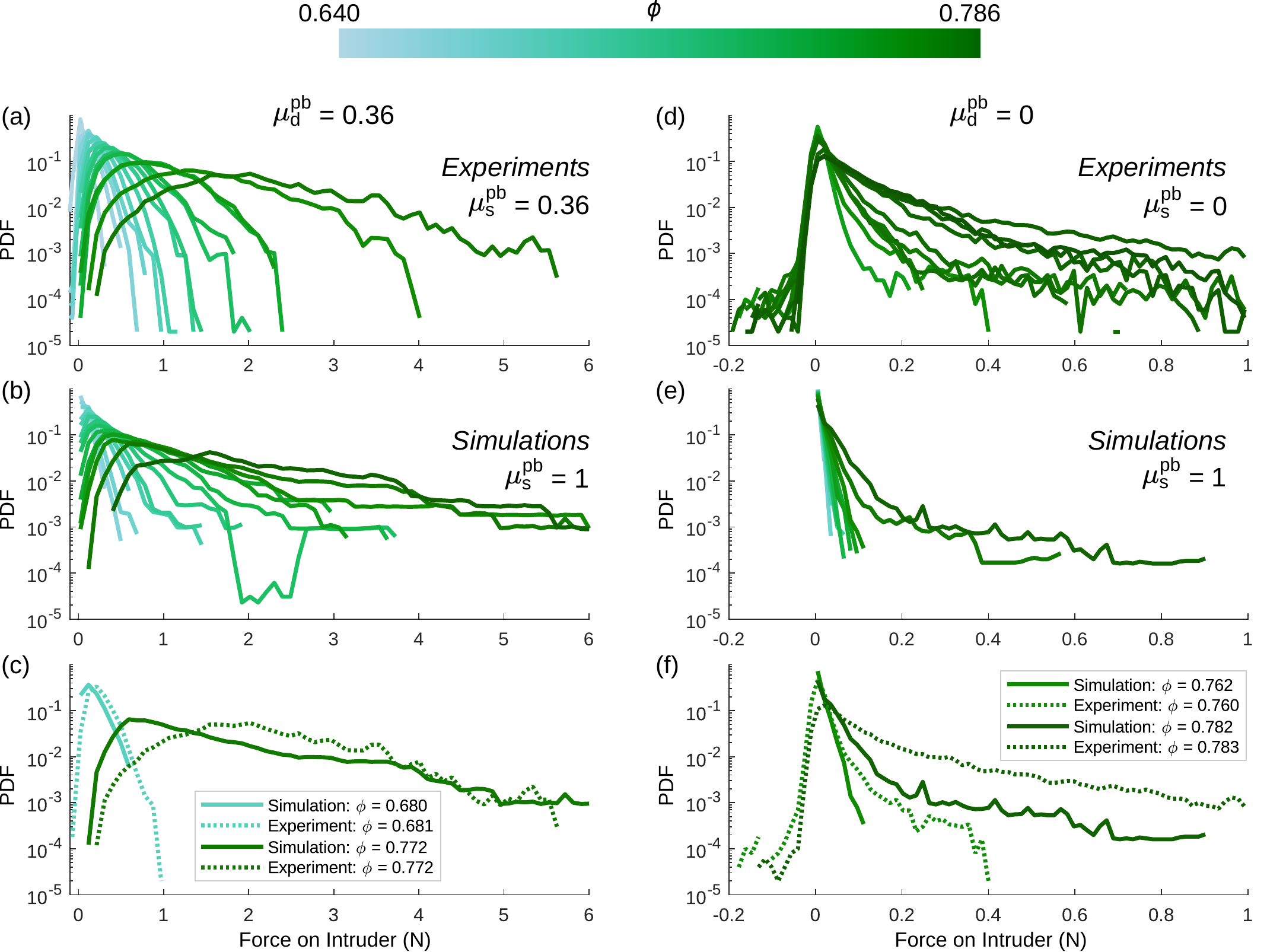}
    \caption{Intruder force distribution for frictional (a-c) and frictionless (d-f) bases. A range of packing fractions have been studied as indicated by the color scale. Experimental (a,d) and simulation (b, e) results show the same trends and range of forces. Panels  (c) and (f) show a direct comparison of the experimental and simulation PDF for two specific values of $\phi$.}
    \label{fig:pdf-force}
\end{figure}

Figures~\ref{fig:pdf-velocity} and~\ref{fig:pdf-force} show the probability distribution functions (PDF) of the intruder velocity and the spring force in simulations and experiments.
We have tested a range of packing fractions, $\phi$, for the frictional and the frictionless base.  The range explored for the frictionless base is narrower because in this case even at relatively high $\phi$ the intruder simply moves at the drive velocity, very rarely getting stuck.  Figure~\ref{fig:pdf-velocity} shows that the agreement between experiments and simulations is remarkably good. Note that the range of velocities observed are consistent as well as the qualitative forms of the distributions.  In the frictional case, for low $\phi$, the velocity distribution has its maximum at the drive speed. However, at high $\phi$, this peak disappears and the maximum occurs at zero velocity, due to the fact that the intruder is stuck most of the time.  These two distinct regimes are separated by a smooth transition region. Following Ref.~\cite{Kozlowski2019}, we refer to these two regimes as a stick--slip regime for high $\phi$ and as an intermittent flow (clogging-like) regime at low $\phi$.   In the case of the frictionless base,
we observe only the intermittent flow, independent of the value of $\phi$. 

Figure~\ref{fig:pdf-force} shows that the spring force PDFs are very similar in experiments and simulations.  The range of forces and the positions of the maxima in the experiments are very well captured by the simulations.  For the frictional base, the PDFs are rather broad with a maximum at non-zero force for high $\phi$. However, for the low $\phi$ (intermittent flow) the PDF has its maximum very near zero force. Negative forces are not present in the simulations due to the modeling of the torque spring as incapable of sustaining tension. In experiments, negative forces are only measured if the spring completely decompresses and the intruder rebounds due to a collision within the central axis (see Ref.~\cite{Kozlowski2019} for more details). For the frictionless base, all $\phi$s lead to PDFs with a maximum at or close to zero force, consistent with the relatively free flow of the intruder. Only at very high $\phi$ do occasional sticking periods lead to some larger forces and therefore longer distribution tails.  

\begin{figure}
    \centering
    \includegraphics[width=1\columnwidth]{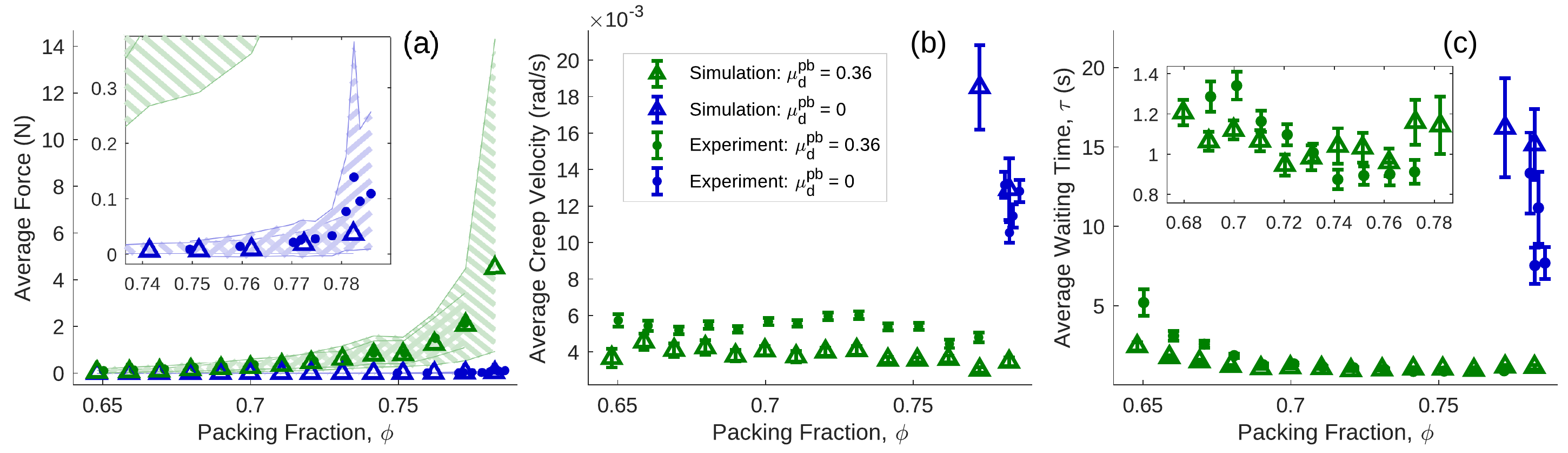}
    \caption{(a) The average force of the force time series as a function of $\phi$, for the frictional (green) and the frictionless (blue) base. Open triangles correspond to simulation data and filled circles to experiments. Each shaded region ranges from the lower 10\% cutoff to the upper 90\% cutoff of the distribution. Error bars indicate the standard error of the average. Inset: Zoom in to highlight the case with no basal friction. (b) The average creep velocity of the intruder during detected sticking periods as a function of $\phi$. (c) The average waiting time as a function of $\phi$. Inset: Zoom in to highlight the case with basal friction.}
    \label{fig:summary-phi}
\end{figure}

Figure~\ref{fig:summary-phi}(a) shows the average force as a function of $\phi$ for both the frictional and the frictionless base. Overall, experimental and simulation values are very similar, with the exception that at high $\phi$ the average forces are higher in simulations, corresponding to the fact that the upper cutoffs in the distributions of Fig.~\ref{fig:pdf-force} are somewhat larger in simulations than in experiments. As we show in Sec.~\ref{sec:basalfriction}, this quantitative difference is a consequence of running these simulations with a higher static basal friction coefficient than that measured in experiments. 

We have measured two additional features of the intruder dynamics: (i) the creep velocity during sticking periods and (ii) the waiting time between consecutive sticking periods. For these statistical quantities, unless noted otherwise, we only present data from runs for which at least 20 sticking periods are detected. During a sticking period, the intruder does not remain fully static but creeps forward as the spring force increases. Figure~\ref{fig:summary-phi}(b) shows that the creep velocity is independent of $\phi$ when basal friction is present and is substantially larger for the frictionless base, both in simulations and experiments. Figure~\ref{fig:summary-phi}(c) shows that the average waiting time is about ten times longer for the frictionless base than for the frictional one at the high packing fractions where sticking events do occur in the frictionless case; sticking events occur less frequently for $\mudpb=0$. With the frictional base, for which sticking events occur at lower
$\phi$, we find a mild increase of the waiting time as $\phi$ is decreased below $0.67$. Here too, simulations and experiments are consistent.

Given the close agreement between experiment and simulation, it appears that our  simulations are a reliable tool for describing, studying, and explaining the dynamics observed in experiments. We next use simulations to explore in more detail the effect of basal friction on the intruder dynamics, which is a difficult task to carry out experimentally.

\subsection{Effect of basal friction}\label{sec:basalfriction}

As we have discussed, there is a dramatic change in the intruder dynamics when friction with the base is removed. This naturally raises the issue of whether the dynamics can be tuned continuously by changing the basal friction, or whether there is a sharp transition at $\mudpb=0$. Controlling basal friction in the experiments is prohibitively difficult, and this is where the simulations provide novel insights.

\begin{figure}
    \centering
    \includegraphics[width=0.75\columnwidth]{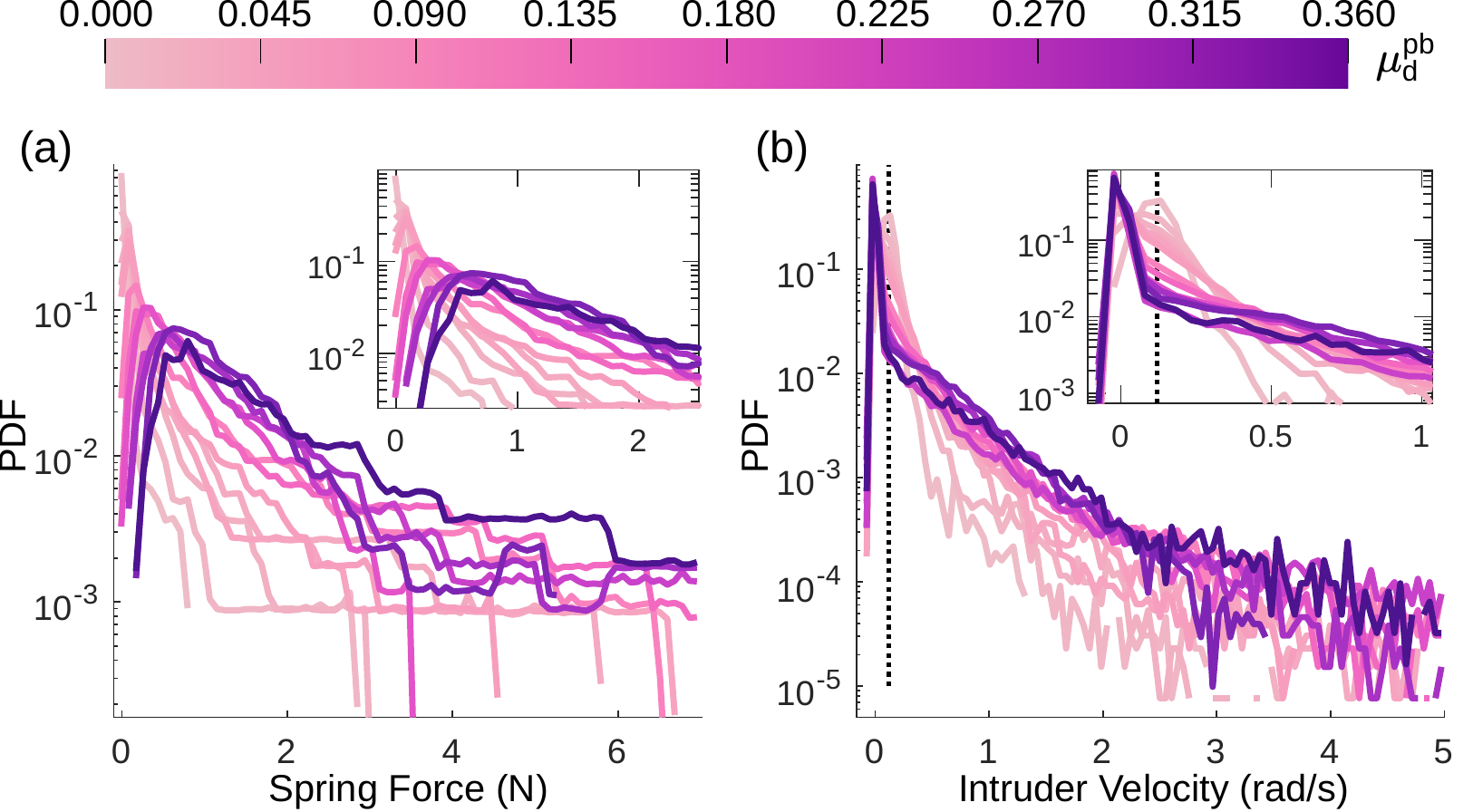}
    \caption{PDF of the spring force (a) and the intruder velocity (b) for a range of basal dynamic friction coefficient ($0 \leq \mudpb \leq 0.36$) and $\muspb=1.0$. Insets: Zoom into the peak values of the PDFs.}
    \label{fig:pdf-dynamic-friction}
\end{figure}

We have run simulations with dynamic friction coefficient $\mudpb$ in the range $[0.0,\, 0.36]$ at a high packing fraction ($\phi=0.7724$) while keeping the static friction coefficient constant at $\muspb=1.0$. 
Figure~\ref{fig:pdf-dynamic-friction} shows the intruder velocity and spring force distributions for a range of $\mudpb$. 
The spring force PDFs shown in Fig.~\ref{fig:pdf-dynamic-friction}(a) reveal that the 
probabilities for large forces within the stick-slip regime decrease
as $\mudpb$ is decreased from 0.36 down to 0.1, indicating that the lowering  of $\mudpb$ (in the stick-slip regime) induces shorter sticking periods. Within the intermittent flow regime ($\mudpb<0.1$), lowering $\mudpb$ leads to the appearance of a sharp peak at zero force, caused mainly by the longer periods of continuous flow between sticks. 
The velocity distribution shown in Fig.~\ref{fig:pdf-dynamic-friction}(b) seems to be nearly independent of $\mudpb$ for $\mudpb > 0.1$, showing a maximum at zero velocity, which is consistent with stick--slip behavior. However, for $\mudpb<0.1$ the maximum of the velocity PDF abruptly shifts to the spring drive speed, a characteristic of the intermittent flow regime. 

\begin{figure}
    \centering
    \includegraphics[width=1\columnwidth]{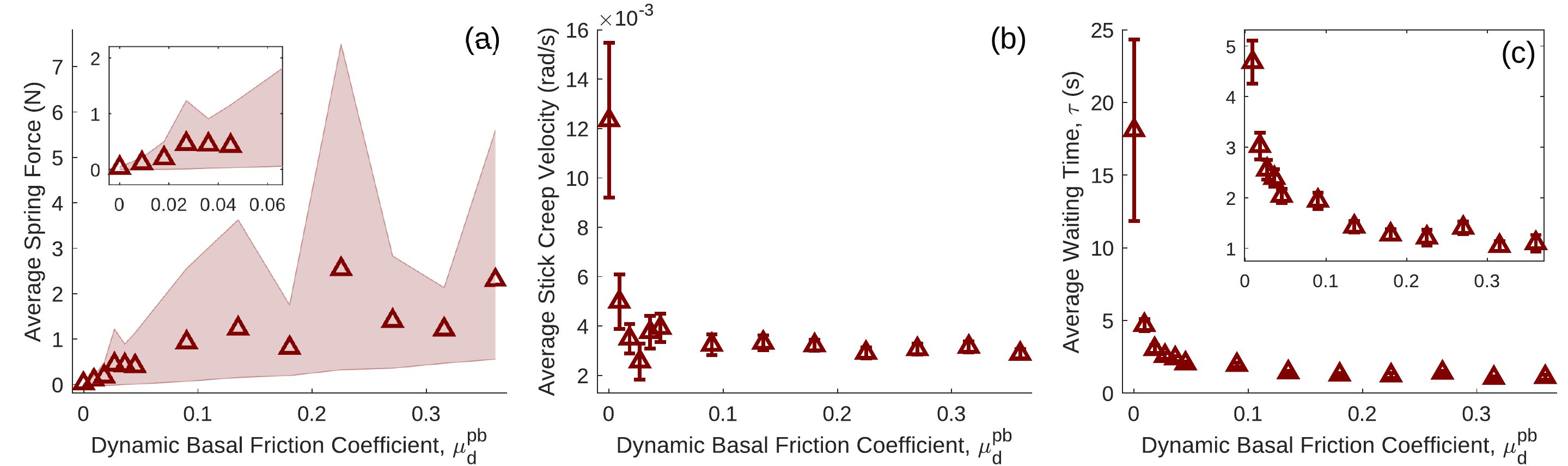}
    \caption{Average force (a), average creep velocity during sticking periods (b), and average waiting time between sticking periods (c) as a function of $\mudpb$ for $\muspb=1.0$. Error bars indicate the standard error of the average and the shaded area indicates the 10\%--90\% percentile in the force distribution. Note that the $\mudpb=0$ data point in (b) and (c) is computed from only 11 events.}
    \label{fig:summary-dynamic-friction}
\end{figure}

Figure~\ref{fig:summary-dynamic-friction} shows
the average force, average creep velocity, and average waiting
time as a function of the dynamic friction coefficient $\mudpb$. Interestingly, the creep velocity and the average waiting time between the end of a sticking period and the beginning of the next one grow very little as the dynamic friction is decreased from $0.36$ to $0.1$. Thus, above $\mudpb \approx 0.1$, a reduction in $\mudpb$ does not significantly change the number of sticking periods (longer waiting times) and stiffness of the packing (larger creep velocities). However, a dramatic increase in both creep velocity and waiting time happens for $\mudpb<0.1$, indicating a rather sharp transition between the stick-slip and intermittent flow regimes.

\begin{figure}
    \centering
    \includegraphics[width=1\columnwidth]{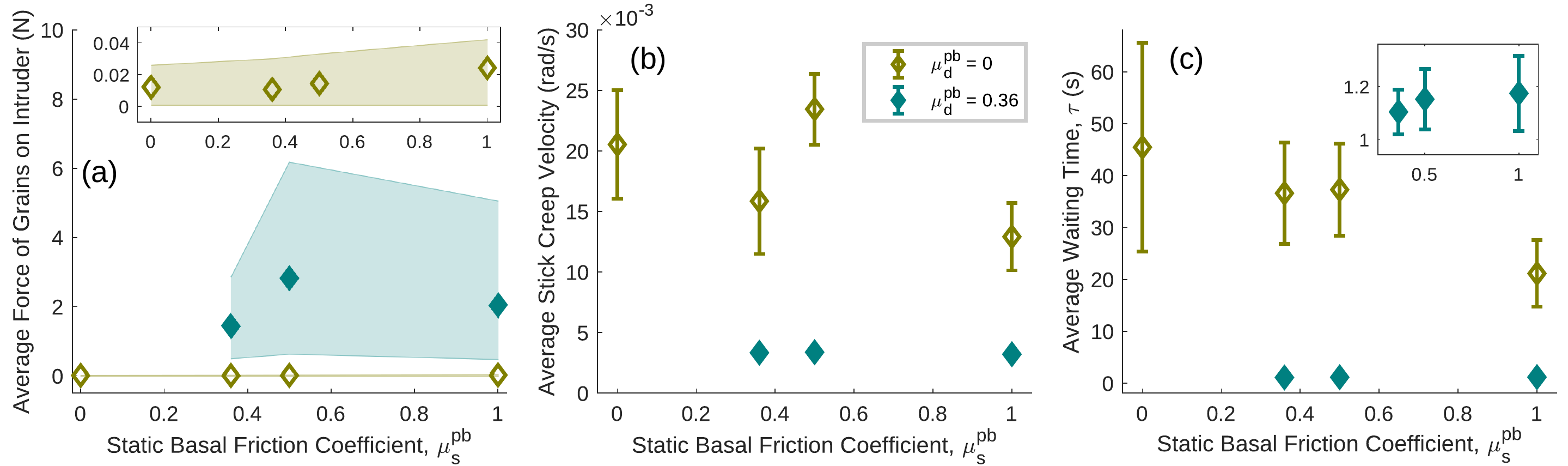}
    \caption{Average force (a), Average creep velocity during sticking periods (b), and average waiting time between sticking periods (c) as a function of the static particle-base friction $\muspb$ for two values of the dynamic friction: $\mudpb =0$ (open diamonds) and $\mudpb=0.36$ (filled diamonds).  Error bars indicate the standard error of the average and the shaded area indicates the 10\%--90\% percentile in the force distribution. Note that all of the $\mudpb=0$ data points are computed from fewer than 20 events. For the filled diamonds, error bars are smaller than the data points.}
    \label{fig:summary-static-friction}
\end{figure}

\begin{figure}
    \centering
    \includegraphics[width=0.9\columnwidth]{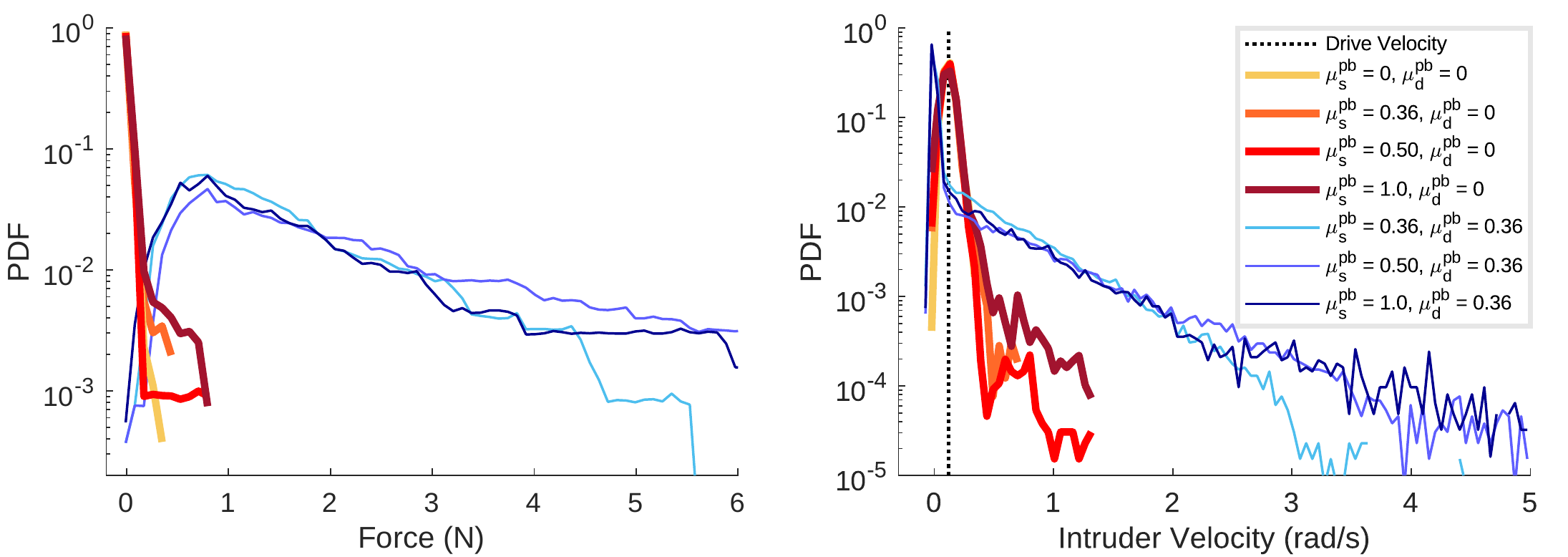}
    \caption{PDF of the spring force (a) and the intruder velocity (b) for a range of basal static friction coefficient ($0.36 \leq \muspb \leq 1.0$) for two values of the dynamic friction coefficient: $\mudpb=0$ (thick lines) and $\mudpb=0.36$ (thin lines). }
    \label{fig:pdf-static-friction}
\end{figure}

Figure~\ref{fig:summary-static-friction} shows the average force, average creep velocity, and average waiting time for different values of basal {\em static} friction coefficient $\muspb$, both for $\mudpb=0$ and $\mudpb=0.36$. In all cases we ensure $\mudpb \leq \muspb$. Static friction seems to play a marginal role in the dynamics. The most salient feature is a marked drop in the spread of the forces (indicated by the width of the shaded region in Fig.~\ref{fig:summary-static-friction}(a)) for $\mudpb = 0.36$  when $\muspb$ is changed from 1.0 to 0.36 . This indicates that $\muspb> \mudpb$ 
 induces the occurrence of some longer lasting sticking periods, without affecting the average force significantly. This finding is confirmed in Fig.~\ref{fig:pdf-static-friction}, which shows  the PDFs for forces and intruder velocities. We note that the PDF in Fig.~\ref{fig:pdf-static-friction}(a) for $\muspb=0.36$ suggests a cutoff at large forces, in agreement with the experimental observation shown in Fig.~\ref{fig:pdf-force}. We recall here that in Fig. \ref{fig:pdf-force} we used $\muspb=1.0$, which is somewhat above the experimental static friction coefficient between the discs and the dry base.  

\section{Conclusions}\label{sec:conclusions}

Inspired by a recent experimental finding on the dynamics of an intruder the size of a single grain dragged though a two-dimensional granular system \cite{Kozlowski2019}, we have developed numerical simulations that allow us to deepen our understanding of  the effect of basal friction. Experimentally, it has been observed that the intruder can flow rather smoothly with occasional short sticks (intermittent flow regime) or show a fully developed stick--slip dynamics. The former is observed if the base on which the particles sit is frictionless, the latter when the base is frictional. Our simulations yield results consistent with these dynamics and are also in fair quantitative agreement with the experiment.

Having the numerical model validated against the experiments, we use the simulations to investigate how the transition from intermittent flow to stick--slip is controlled by both the dynamic and the static friction coefficient of the particle--base interaction. 
We have shown that the transition between the two dynamic regimes (intermittent flow and stick-slip) is clearly controlled by the dynamic friction coefficient with the base and not by the static friction coefficient. One may intuitively expect that static friction would play a major role by contributing to the stability of jammed configurations. However, one has to bear in mind that jammed configurations are reached through a dynamic process. The development of rigidity is a result of the interactions at play during motion of the particles, making $\mudpb$ the more relevant factor in determining the waiting times between sticking events. Our interpretation is that, for $\mudpb>0.1$, particle motion is damped strongly enough to allow rapid formation of stable, static force network structures. Further increase in $\mudpb$ can only marginally reduce these waiting times. Decreasing $\mudpb$ below $0.1$, however, reduces the damping enough so that particles do not quickly come to rest after a slip event is initiated.  Particle--particle dissipative interactions (inelasticity and friction) then play the dominant role in slowing particles down, and this leads only to occasional clogging rather than strong sticking periods.

Static friction is expected to affect the stability of the jammed states once they occur. One of the 
affected features is the decay of the intruder force PDF for large forces, which is associated with the stability of the jammed states. The larger $\muspb$ is,  the higher the probability of finding strong intruder forces. This is indeed confirmed in Fig.~\ref{fig:pdf-static-friction}, both for $\mudpb=0$ and $\mudpb=0.36$.  Increasing $\muspb$ does not, however, induce a change of 
dynamic regime (intermittent flow or stick--slip); the main peaks in the force and velocity distributions remain unaltered.
In short, static friction can only help the stick periods last longer when they do occur. The initiation of these sticking events, however, is determined solely by the basal dynamic friction. 

Our results show that the additional background dissipation (especially dynamic friction) provided by the supporting substrate is essential for observing a clear stick--slip dynamics in the case of single-grain perturbations. However, stick--slip is observed without the presence of a frictional substrate if a driving force is applied over a length scale that is much larger than the characteristic particle size~\cite{SimGranularSeismicPicaCiamarra2011,stickslipalbert,stickslippulldiskoutMetayer,stickslipnasuno,stickslipcracklingagheal,SimGranularSeismicPicaCiamarra2011}. One open question is whether an additional background dissipation in the case of
larger perturbations can alter the dynamics. Existing studies on granular systems immersed in viscous fluids (which provide an additional dissipation) indeed do show a strong effect on the stick--slip dynamics for large scale perturbations~\cite{Higashi2009}. However, the effects observed seem to be more connected to lubrication than to additional ``background'' dissipation. A natural question is then how parameters of an immersed system, such as particle size and fluid viscosity, can possibly strengthen or weaken stick-slip behavior as lubrication and drag compete, and whether the effects of fluid drag might be similar to those due to frictional dissipation.  

Finally, we note that simulations provide a much higher resolution for velocities and forces than those achieved in the experiments. In order to compare results we have used in the simulation data analysis the same velocity threshold to detect sticking periods as was used in analyzing the experimental data. This overlooks a number of detailed features observed in the simulations during the sticking periods that may also be present in the experiments but were not resolvable in the experimental system.  In particular, the creep observed in simulations has a very rich structure of \emph{micro-slips}. With the simulation model now validated, it will be interesting to explore the details of the dynamics that are difficult to measure in experiments, including the interparticle forces and particle displacements during micro-slip events.



\begin{acknowledgments}
We thank Karen Daniels for valuable discussions. This work was supported by the US Army Research Office through grant W911NF1810184 and by the Keck Foundation.  L. A. P. and C. M. C. acknowledge support by Universidad Tecnol\'ogica Nacional through grants PID-MAUTNLP0004415 and PID-MAIFIBA0004434TC and CONICET through grant RES-1225-17. C. M. C. also thanks the Norma Hoermann Foundation for partial funding for his visit to NJ. L. K. was supported in part by NSF Grant No. 1521717. 
\end{acknowledgments}
\bibliography{granulates}

\begin{thebibliography}{23}%
\makeatletter
\providecommand \@ifxundefined [1]{%
 \@ifx{#1\undefined}
}%
\providecommand \@ifnum [1]{%
 \ifnum #1\expandafter \@firstoftwo
 \else \expandafter \@secondoftwo
 \fi
}%
\providecommand \@ifx [1]{%
 \ifx #1\expandafter \@firstoftwo
 \else \expandafter \@secondoftwo
 \fi
}%
\providecommand \natexlab [1]{#1}%
\providecommand \enquote  [1]{``#1''}%
\providecommand \bibnamefont  [1]{#1}%
\providecommand \bibfnamefont [1]{#1}%
\providecommand \citenamefont [1]{#1}%
\providecommand \href@noop [0]{\@secondoftwo}%
\providecommand \href [0]{\begingroup \@sanitize@url \@href}%
\providecommand \@href[1]{\@@startlink{#1}\@@href}%
\providecommand \@@href[1]{\endgroup#1\@@endlink}%
\providecommand \@sanitize@url [0]{\catcode `\\12\catcode `\$12\catcode
  `\&12\catcode `\#12\catcode `\^12\catcode `\_12\catcode `\%12\relax}%
\providecommand \@@startlink[1]{}%
\providecommand \@@endlink[0]{}%
\providecommand \url  [0]{\begingroup\@sanitize@url \@url }%
\providecommand \@url [1]{\endgroup\@href {#1}{\urlprefix }}%
\providecommand \urlprefix  [0]{URL }%
\providecommand \Eprint [0]{\href }%
\providecommand \doibase [0]{http://dx.doi.org/}%
\providecommand \selectlanguage [0]{\@gobble}%
\providecommand \bibinfo  [0]{\@secondoftwo}%
\providecommand \bibfield  [0]{\@secondoftwo}%
\providecommand \translation [1]{[#1]}%
\providecommand \BibitemOpen [0]{}%
\providecommand \bibitemStop [0]{}%
\providecommand \bibitemNoStop [0]{.\EOS\space}%
\providecommand \EOS [0]{\spacefactor3000\relax}%
\providecommand \BibitemShut  [1]{\csname bibitem#1\endcsname}%
\let\auto@bib@innerbib\@empty
\bibitem [{\citenamefont {Hayman}\ \emph {et~al.}(2011)\citenamefont {Hayman},
  \citenamefont {Duclou{\'e}}, \citenamefont {Foco},\ and\ \citenamefont
  {Daniels}}]{granularfaultdanielshayman}%
  \BibitemOpen
  \bibfield  {author} {\bibinfo {author} {\bibfnamefont {N.~W.}\ \bibnamefont
  {Hayman}}, \bibinfo {author} {\bibfnamefont {L.}~\bibnamefont {Duclou{\'e}}},
  \bibinfo {author} {\bibfnamefont {K.~L.}\ \bibnamefont {Foco}}, \ and\
  \bibinfo {author} {\bibfnamefont {K.~E.}\ \bibnamefont {Daniels}},\ }\href
  {\doibase 10.1007/s00024-011-0269-3} {\bibfield  {journal} {\bibinfo
  {journal} {Pure Appl. Geophys.}\ }\textbf {\bibinfo {volume} {168}},\
  \bibinfo {pages} {2239} (\bibinfo {year} {2011})}\BibitemShut {NoStop}%
\bibitem [{\citenamefont {Bar\'es}\ \emph {et~al.}(2017)\citenamefont
  {Bar\'es}, \citenamefont {Wang}, \citenamefont {Wang}, \citenamefont
  {Bertrand}, \citenamefont {O'Hern},\ and\ \citenamefont
  {Behringer}}]{localglobalavalanchesbares}%
  \BibitemOpen
  \bibfield  {author} {\bibinfo {author} {\bibfnamefont {J.}~\bibnamefont
  {Bar\'es}}, \bibinfo {author} {\bibfnamefont {D.}~\bibnamefont {Wang}},
  \bibinfo {author} {\bibfnamefont {D.}~\bibnamefont {Wang}}, \bibinfo {author}
  {\bibfnamefont {T.}~\bibnamefont {Bertrand}}, \bibinfo {author}
  {\bibfnamefont {C.~S.}\ \bibnamefont {O'Hern}}, \ and\ \bibinfo {author}
  {\bibfnamefont {R.~P.}\ \bibnamefont {Behringer}},\ }\href {\doibase
  10.1103/PhysRevE.96.052902} {\bibfield  {journal} {\bibinfo  {journal} {Phys.
  Rev. E}\ }\textbf {\bibinfo {volume} {96}},\ \bibinfo {pages} {052902}
  (\bibinfo {year} {2017})}\BibitemShut {NoStop}%
\bibitem [{\citenamefont {{Pica Ciamarra}}\ \emph {et~al.}(2011)\citenamefont
  {{Pica Ciamarra}}, \citenamefont {Lippiello}, \citenamefont {de~Arcangelis},\
  and\ \citenamefont {Godano}}]{SimGranularSeismicPicaCiamarra2011}%
  \BibitemOpen
  \bibfield  {author} {\bibinfo {author} {\bibfnamefont {M.}~\bibnamefont
  {{Pica Ciamarra}}}, \bibinfo {author} {\bibfnamefont {E.}~\bibnamefont
  {Lippiello}}, \bibinfo {author} {\bibfnamefont {L.}~\bibnamefont
  {de~Arcangelis}}, \ and\ \bibinfo {author} {\bibfnamefont {C.}~\bibnamefont
  {Godano}},\ }\href {\doibase 10.1209/0295-5075/95/54002} {\bibfield
  {journal} {\bibinfo  {journal} {Europhys. Lett.}\ }\textbf {\bibinfo {volume}
  {95}},\ \bibinfo {pages} {54002} (\bibinfo {year} {2011})}\BibitemShut
  {NoStop}%
\bibitem [{\citenamefont {Albert}\ \emph {et~al.}(2001)\citenamefont {Albert},
  \citenamefont {Tegzes}, \citenamefont {Albert}, \citenamefont {Sample},
  \citenamefont {Barab\'asi}, \citenamefont {Vicsek}, \citenamefont {Kahng},\
  and\ \citenamefont {Schiffer}}]{stickslipalbert}%
  \BibitemOpen
  \bibfield  {author} {\bibinfo {author} {\bibfnamefont {I.}~\bibnamefont
  {Albert}}, \bibinfo {author} {\bibfnamefont {P.}~\bibnamefont {Tegzes}},
  \bibinfo {author} {\bibfnamefont {R.}~\bibnamefont {Albert}}, \bibinfo
  {author} {\bibfnamefont {J.~G.}\ \bibnamefont {Sample}}, \bibinfo {author}
  {\bibfnamefont {A.~L.}\ \bibnamefont {Barab\'asi}}, \bibinfo {author}
  {\bibfnamefont {T.}~\bibnamefont {Vicsek}}, \bibinfo {author} {\bibfnamefont
  {B.}~\bibnamefont {Kahng}}, \ and\ \bibinfo {author} {\bibfnamefont
  {P.}~\bibnamefont {Schiffer}},\ }\href {\doibase 10.1103/PhysRevE.64.031307}
  {\bibfield  {journal} {\bibinfo  {journal} {Phys. Rev. E}\ }\textbf {\bibinfo
  {volume} {64}},\ \bibinfo {pages} {031307} (\bibinfo {year}
  {2001})}\BibitemShut {NoStop}%
\bibitem [{\citenamefont {M{\'{e}}tayer}\ \emph {et~al.}(2011)\citenamefont
  {M{\'{e}}tayer}, \citenamefont {{Suntrup III}}, \citenamefont {Radin},
  \citenamefont {Swinney},\ and\ \citenamefont
  {Schr{\"{o}}ter}}]{stickslippulldiskoutMetayer}%
  \BibitemOpen
  \bibfield  {author} {\bibinfo {author} {\bibfnamefont {J.-F.}\ \bibnamefont
  {M{\'{e}}tayer}}, \bibinfo {author} {\bibfnamefont {D.~J.}\ \bibnamefont
  {{Suntrup III}}}, \bibinfo {author} {\bibfnamefont {C.}~\bibnamefont
  {Radin}}, \bibinfo {author} {\bibfnamefont {H.~L.}\ \bibnamefont {Swinney}},
  \ and\ \bibinfo {author} {\bibfnamefont {M.}~\bibnamefont {Schr{\"{o}}ter}},\
  }\href {\doibase 10.1209/0295-5075/93/64003} {\bibfield  {journal} {\bibinfo
  {journal} {Europhys. Lett.}\ }\textbf {\bibinfo {volume} {93}},\ \bibinfo
  {pages} {64003} (\bibinfo {year} {2011})}\BibitemShut {NoStop}%
\bibitem [{\citenamefont {Nasuno}\ \emph {et~al.}(1998)\citenamefont {Nasuno},
  \citenamefont {Kudrolli}, \citenamefont {Bak},\ and\ \citenamefont
  {Gollub}}]{stickslipnasuno}%
  \BibitemOpen
  \bibfield  {author} {\bibinfo {author} {\bibfnamefont {S.}~\bibnamefont
  {Nasuno}}, \bibinfo {author} {\bibfnamefont {A.}~\bibnamefont {Kudrolli}},
  \bibinfo {author} {\bibfnamefont {A.}~\bibnamefont {Bak}}, \ and\ \bibinfo
  {author} {\bibfnamefont {J.~P.}\ \bibnamefont {Gollub}},\ }\href {\doibase
  10.1103/PhysRevE.58.2161} {\bibfield  {journal} {\bibinfo  {journal} {Phys.
  Rev. E}\ }\textbf {\bibinfo {volume} {58}},\ \bibinfo {pages} {2161}
  (\bibinfo {year} {1998})}\BibitemShut {NoStop}%
\bibitem [{\citenamefont {{Abed Zadeh}}\ \emph {et~al.}(2019)\citenamefont
  {{Abed Zadeh}}, \citenamefont {{Bar{\'e}s}},\ and\ \citenamefont
  {{Behringer}}}]{stickslipcracklingagheal}%
  \BibitemOpen
  \bibfield  {author} {\bibinfo {author} {\bibfnamefont {A.}~\bibnamefont
  {{Abed Zadeh}}}, \bibinfo {author} {\bibfnamefont {J.}~\bibnamefont
  {{Bar{\'e}s}}}, \ and\ \bibinfo {author} {\bibfnamefont {R.~P.}\ \bibnamefont
  {{Behringer}}},\ }\href {\doibase https://doi.org/10.1103/PhysRevE.99.040901}
  {\bibfield  {journal} {\bibinfo  {journal} {Phys. Rev. E}\ }\textbf {\bibinfo
  {volume} {99}},\ \bibinfo {pages} {040901(R)} (\bibinfo {year}
  {2019})}\BibitemShut {NoStop}%
\bibitem [{\citenamefont {Jaeger}\ \emph {et~al.}(1996)\citenamefont {Jaeger},
  \citenamefont {Nagel},\ and\ \citenamefont {Behringer}}]{jaeger96b}%
  \BibitemOpen
  \bibfield  {author} {\bibinfo {author} {\bibfnamefont {H.~M.}\ \bibnamefont
  {Jaeger}}, \bibinfo {author} {\bibfnamefont {S.~R.}\ \bibnamefont {Nagel}}, \
  and\ \bibinfo {author} {\bibfnamefont {R.~P.}\ \bibnamefont {Behringer}},\
  }\href@noop {} {\bibfield  {journal} {\bibinfo  {journal} {Reviews of Modern
  Physics}\ }\textbf {\bibinfo {volume} {68}},\ \bibinfo {pages} {1259}
  (\bibinfo {year} {1996})}\BibitemShut {NoStop}%
\bibitem [{\citenamefont {Olson~Reichhardt}\ and\ \citenamefont
  {Reichhardt}(2010)}]{probeintruderreichhardt}%
  \BibitemOpen
  \bibfield  {author} {\bibinfo {author} {\bibfnamefont {C.~J.}\ \bibnamefont
  {Olson~Reichhardt}}\ and\ \bibinfo {author} {\bibfnamefont {C.}~\bibnamefont
  {Reichhardt}},\ }\href {\doibase 10.1103/PhysRevE.82.051306} {\bibfield
  {journal} {\bibinfo  {journal} {Phys. Rev. E}\ }\textbf {\bibinfo {volume}
  {82}},\ \bibinfo {pages} {051306} (\bibinfo {year} {2010})}\BibitemShut
  {NoStop}%
\bibitem [{\citenamefont {Candelier}\ and\ \citenamefont
  {Dauchot}(2010)}]{intrudervibrationgdauchot}%
  \BibitemOpen
  \bibfield  {author} {\bibinfo {author} {\bibfnamefont {R.}~\bibnamefont
  {Candelier}}\ and\ \bibinfo {author} {\bibfnamefont {O.}~\bibnamefont
  {Dauchot}},\ }\href {\doibase 10.1103/PhysRevE.81.011304} {\bibfield
  {journal} {\bibinfo  {journal} {Phys. Rev. E}\ }\textbf {\bibinfo {volume}
  {81}},\ \bibinfo {pages} {011304} (\bibinfo {year} {2010})}\BibitemShut
  {NoStop}%
\bibitem [{\citenamefont {Kolb}\ \emph {et~al.}(2013)\citenamefont {Kolb},
  \citenamefont {Cixous}, \citenamefont {Gaudouen},\ and\ \citenamefont
  {Darnige}}]{dragforcecavityformationintruderkolb}%
  \BibitemOpen
  \bibfield  {author} {\bibinfo {author} {\bibfnamefont {E.}~\bibnamefont
  {Kolb}}, \bibinfo {author} {\bibfnamefont {P.}~\bibnamefont {Cixous}},
  \bibinfo {author} {\bibfnamefont {N.}~\bibnamefont {Gaudouen}}, \ and\
  \bibinfo {author} {\bibfnamefont {T.}~\bibnamefont {Darnige}},\ }\href
  {\doibase 10.1103/PhysRevE.87.032207} {\bibfield  {journal} {\bibinfo
  {journal} {Phys. Rev. E}\ }\textbf {\bibinfo {volume} {87}},\ \bibinfo
  {pages} {032207} (\bibinfo {year} {2013})}\BibitemShut {NoStop}%
\bibitem [{\citenamefont {Geng}\ and\ \citenamefont
  {Behringer}(2005)}]{slowdraggeng}%
  \BibitemOpen
  \bibfield  {author} {\bibinfo {author} {\bibfnamefont {J.}~\bibnamefont
  {Geng}}\ and\ \bibinfo {author} {\bibfnamefont {R.~P.}\ \bibnamefont
  {Behringer}},\ }\href {\doibase 10.1103/PhysRevE.71.011302} {\bibfield
  {journal} {\bibinfo  {journal} {Phys. Rev. E}\ }\textbf {\bibinfo {volume}
  {71}},\ \bibinfo {pages} {011302} (\bibinfo {year} {2005})}\BibitemShut
  {NoStop}%
\bibitem [{\citenamefont {Seguin}\ \emph {et~al.}(2016)\citenamefont {Seguin},
  \citenamefont {Coulais}, \citenamefont {Martinez}, \citenamefont {Bertho},\
  and\ \citenamefont {Gondret}}]{rheologyintruderexperimentseguin}%
  \BibitemOpen
  \bibfield  {author} {\bibinfo {author} {\bibfnamefont {A.}~\bibnamefont
  {Seguin}}, \bibinfo {author} {\bibfnamefont {C.}~\bibnamefont {Coulais}},
  \bibinfo {author} {\bibfnamefont {F.}~\bibnamefont {Martinez}}, \bibinfo
  {author} {\bibfnamefont {Y.}~\bibnamefont {Bertho}}, \ and\ \bibinfo {author}
  {\bibfnamefont {P.}~\bibnamefont {Gondret}},\ }\href {\doibase
  10.1103/PhysRevE.93.012904} {\bibfield  {journal} {\bibinfo  {journal} {Phys.
  Rev. E}\ }\textbf {\bibinfo {volume} {93}},\ \bibinfo {pages} {012904}
  (\bibinfo {year} {2016})}\BibitemShut {NoStop}%
\bibitem [{\citenamefont {Tordesillas}\ \emph {et~al.}(2014)\citenamefont
  {Tordesillas}, \citenamefont {Hilton},\ and\ \citenamefont
  {Tobin}}]{sticksliptordesillas}%
  \BibitemOpen
  \bibfield  {author} {\bibinfo {author} {\bibfnamefont {A.}~\bibnamefont
  {Tordesillas}}, \bibinfo {author} {\bibfnamefont {J.~E.}\ \bibnamefont
  {Hilton}}, \ and\ \bibinfo {author} {\bibfnamefont {S.~T.}\ \bibnamefont
  {Tobin}},\ }\href {\doibase 10.1103/PhysRevE.89.042207} {\bibfield  {journal}
  {\bibinfo  {journal} {Phys. Rev. E}\ }\textbf {\bibinfo {volume} {89}},\
  \bibinfo {pages} {042207} (\bibinfo {year} {2014})}\BibitemShut {NoStop}%
\bibitem [{\citenamefont {Kozlowski}\ \emph {et~al.}(2019)\citenamefont
  {Kozlowski}, \citenamefont {Carlevaro}, \citenamefont {Daniels},
  \citenamefont {Kondic}, \citenamefont {Pugnaloni}, \citenamefont {Socolar},
  \citenamefont {Zheng},\ and\ \citenamefont {Behringer}}]{Kozlowski2019}%
  \BibitemOpen
  \bibfield  {author} {\bibinfo {author} {\bibfnamefont {R.}~\bibnamefont
  {Kozlowski}}, \bibinfo {author} {\bibfnamefont {C.~M.}\ \bibnamefont
  {Carlevaro}}, \bibinfo {author} {\bibfnamefont {K.~E.}\ \bibnamefont
  {Daniels}}, \bibinfo {author} {\bibfnamefont {L.}~\bibnamefont {Kondic}},
  \bibinfo {author} {\bibfnamefont {L.~A.}\ \bibnamefont {Pugnaloni}}, \bibinfo
  {author} {\bibfnamefont {J.~E.~S.}\ \bibnamefont {Socolar}}, \bibinfo
  {author} {\bibfnamefont {H.}~\bibnamefont {Zheng}}, \ and\ \bibinfo {author}
  {\bibfnamefont {R.~P.}\ \bibnamefont {Behringer}},\ }\href {\doibase
  10.1103/PhysRevE.100.032905} {\bibfield  {journal} {\bibinfo  {journal}
  {Phys. Rev. E}\ }\textbf {\bibinfo {volume} {100}},\ \bibinfo {pages}
  {032905} (\bibinfo {year} {2019})}\BibitemShut {NoStop}%
\bibitem [{\citenamefont {Zheng}\ \emph {et~al.}(2014)\citenamefont {Zheng},
  \citenamefont {Dijksman},\ and\ \citenamefont {Behringer}}]{shearjamnobfhu}%
  \BibitemOpen
  \bibfield  {author} {\bibinfo {author} {\bibfnamefont {H.}~\bibnamefont
  {Zheng}}, \bibinfo {author} {\bibfnamefont {J.~A.}\ \bibnamefont {Dijksman}},
  \ and\ \bibinfo {author} {\bibfnamefont {R.~P.}\ \bibnamefont {Behringer}},\
  }\href {http://stacks.iop.org/0295-5075/107/i=3/a=34005} {\bibfield
  {journal} {\bibinfo  {journal} {Europhys. Lett.}\ }\textbf {\bibinfo {volume}
  {107}},\ \bibinfo {pages} {34005} (\bibinfo {year} {2014})}\BibitemShut
  {NoStop}%
\bibitem [{box()}]{box2d}%
  \BibitemOpen
  \href@noop {} {\enquote {\bibinfo {title} {Box2d physics engine},}\ }\bibinfo
  {note} {Available at http://www.box2d.org}\BibitemShut {NoStop}%
\bibitem [{\citenamefont {Catto}(2005)}]{catto}%
  \BibitemOpen
  \bibfield  {author} {\bibinfo {author} {\bibfnamefont {E.}~\bibnamefont
  {Catto}},\ }\href@noop {} {\enquote {\bibinfo {title} {Iterative dynamics
  with temporal coherence},}\ } (\bibinfo {year} {2005}),\ \bibinfo {note}
  {available at http://box2d.googlecode.com/}\BibitemShut {NoStop}%
\bibitem [{\citenamefont {Pytlos}\ \emph {et~al.}(2015)\citenamefont {Pytlos},
  \citenamefont {Gilbert},\ and\ \citenamefont {Smith}}]{pytlos2015modelling}%
  \BibitemOpen
  \bibfield  {author} {\bibinfo {author} {\bibfnamefont {M.}~\bibnamefont
  {Pytlos}}, \bibinfo {author} {\bibfnamefont {M.}~\bibnamefont {Gilbert}}, \
  and\ \bibinfo {author} {\bibfnamefont {C.~C.}\ \bibnamefont {Smith}},\ }\href
  {\doibase 10.1680/jgele.15.00067} {\bibfield  {journal} {\bibinfo  {journal}
  {Geotech. Lett.}\ }\textbf {\bibinfo {volume} {5}},\ \bibinfo {pages} {243}
  (\bibinfo {year} {2015})}\BibitemShut {NoStop}%
\bibitem [{\citenamefont {Carlevaro}\ and\ \citenamefont
  {Pugnaloni}(2011)}]{carlevaro_jsm11}%
  \BibitemOpen
  \bibfield  {author} {\bibinfo {author} {\bibfnamefont {C.}~\bibnamefont
  {Carlevaro}}\ and\ \bibinfo {author} {\bibfnamefont {L.}~\bibnamefont
  {Pugnaloni}},\ }\href {\doibase 10.1088/1742-5468/2011/01/P01007} {\bibfield
  {journal} {\bibinfo  {journal} {J. Stat. Mech.}\ }\textbf {\bibinfo {volume}
  {11}},\ \bibinfo {pages} {01007} (\bibinfo {year} {2011})}\BibitemShut
  {NoStop}%
\bibitem [{\citenamefont {Irastorza}\ \emph {et~al.}(2013)\citenamefont
  {Irastorza}, \citenamefont {Carlevaro},\ and\ \citenamefont
  {Pugnaloni}}]{irastorza2013exact}%
  \BibitemOpen
  \bibfield  {author} {\bibinfo {author} {\bibfnamefont {R.~M.}\ \bibnamefont
  {Irastorza}}, \bibinfo {author} {\bibfnamefont {C.~M.}\ \bibnamefont
  {Carlevaro}}, \ and\ \bibinfo {author} {\bibfnamefont {L.~A.}\ \bibnamefont
  {Pugnaloni}},\ }\href {\doibase 10.1088/1742-5468/2013/12/P12012} {\bibfield
  {journal} {\bibinfo  {journal} {J. Stat. Mech.}\ }\textbf {\bibinfo {volume}
  {2013}},\ \bibinfo {pages} {P12012} (\bibinfo {year} {2013})}\BibitemShut
  {NoStop}%
\bibitem [{\citenamefont {S{\'a}nchez}\ \emph {et~al.}(2014)\citenamefont
  {S{\'a}nchez}, \citenamefont {Carlevaro},\ and\ \citenamefont
  {Pugnaloni}}]{sanchez2014effect}%
  \BibitemOpen
  \bibfield  {author} {\bibinfo {author} {\bibfnamefont {M.}~\bibnamefont
  {S{\'a}nchez}}, \bibinfo {author} {\bibfnamefont {C.~M.}\ \bibnamefont
  {Carlevaro}}, \ and\ \bibinfo {author} {\bibfnamefont {L.~A.}\ \bibnamefont
  {Pugnaloni}},\ }\href {\doibase 10.1177/1077546313480544} {\bibfield
  {journal} {\bibinfo  {journal} {J. Vibration Control}\ }\textbf {\bibinfo
  {volume} {20}},\ \bibinfo {pages} {1846} (\bibinfo {year}
  {2014})}\BibitemShut {NoStop}%
\bibitem [{\citenamefont {Higashi}\ and\ \citenamefont
  {Sumita}(2009)}]{Higashi2009}%
  \BibitemOpen
  \bibfield  {author} {\bibinfo {author} {\bibfnamefont {N.}~\bibnamefont
  {Higashi}}\ and\ \bibinfo {author} {\bibfnamefont {I.}~\bibnamefont
  {Sumita}},\ }\href {\doibase 10.1029/2008JB005999} {\bibfield  {journal}
  {\bibinfo  {journal} {Journal of Geophysical Research: Solid Earth}\ }\textbf
  {\bibinfo {volume} {114}},\ \bibinfo {pages} {B04413} (\bibinfo {year}
  {2009})}\BibitemShut {NoStop}%
\end{thebibliography}%

\end{document}